\documentclass[twocolumn]{aastex631}

\begin{document}

\title{Galaxies at $z\gtrsim 10$: $\Lambda$CDM predicts increased Star Formation Efficiency}
\

\correspondingauthor{Francisco Prada}
\email{f.prada@csic.es}

\author[0000-0001-7145-8674]{Francisco Prada}
\affiliation{Instituto de Astrof\'isica de Andaluc\'ia (IAA-CSIC), Granada, E-18008, Spain}

\author{Peter Behroozi}
\affiliation{Department of Astronomy and Steward Observatory, University of Arizona, Tucson, AZ 85721, USA}

\author[0000-0002-5316-9171]{Tomoaki Ishiyama}
\affiliation{Digital Transformation Enhancement Council, Chiba University, 1-33, Yayoi-cho, Inage-ku, Chiba, 263-8522, Japan}

\author[0000-0001-9737-4559]{Enrique Pérez}
\affiliation{Instituto de Astrof\'isica de Andaluc\'ia (IAA-CSIC), Granada, E-18008, Spain}

\author{Anatoly Klypin}
\affiliation{Department of Astronomy, University of Virginia, Charlottesville, VA 22904, USA}

\author[0000-0002-9373-3865]{Xin Wang}
\affil{School of Astronomy and Space Science, University of Chinese Academy of Sciences (UCAS), Beijing 100049, China}
\affil{Institute for Frontiers in Astronomy and Astrophysics, Beijing Normal University,  Beijing 102206, China}
\affil{National Astronomical Observatories, Chinese Academy of Sciences, Beijing 100101, China}

\author{José Ruedas}
\affiliation{Instituto de Astrof\'isica de Andaluc\'ia (IAA-CSIC), Granada, E-18008, Spain}

\begin{abstract}

We show that the rest-frame UV statistics and global properties of galaxies at $7 \lesssim z \lesssim 14$ are naturally reproduced within the standard $\Lambda$CDM framework when galaxy formation is modeled with \textsc{UniverseMachine} applied to the high-resolution \textsc{Uchuu} $N$-body simulation. Our model matches the UV luminosity functions over five magnitudes and reproduces the evolution of the UV (and inferred star formation rate) density once internal dust attenuation is included. Comparisons with spectroscopically confirmed JWST/HST galaxies show good agreement with the stellar mass--SFR and stellar mass--UV luminosity relations. In contrast, earlier claims of insufficient stellar masses at $z \sim 8$ are inconsistent with our model and are likely driven by systematic uncertainties, including AGN contamination, dust attenuation, and the lack of JWST/MIRI constraints. A key prediction is that the star-formation efficiency increases with redshift at fixed halo mass, reaching $\sim$2--3\% of baryons converted into stars by $z \sim 10$--12. These results demonstrate that current JWST observations of early galaxy populations can be explained within the $\Lambda$CDM framework.

\end{abstract}




\section{Introduction} \label{sec:intro}

In their study, \cite{Labbe2023} used multi-band infrared images captured by the James Webb Space Telescope (JWST) to discover a population of red massive galaxies that formed approximately 600 million years after the Big Bang. The authors reported an extraordinarily large density of these galaxies, with stellar masses exceeding $10^{10}\,M_{\odot}$, which, if confirmed, challenges the standard cosmological model as suggested by recent studies \citep{Boylan-Kolchin2023, Lovell2023, liu.2026}. 

However, this conclusion is disputed. In his letter, we argue that during the early epochs of the universe, the stellar mass-to-light ratio could not have reached the values reported by \cite{Labbe2023}. A galaxy formation model based on our 2.1 trillion particle \textsc{Uchuu} simulation \citep{Ishiyama2021}, within the standard $\Lambda$CDM cosmology, supports this hypothesis. The model predicts the formation of massive galaxies with higher ultraviolet (UV) luminosities, forming stars at rates of several hundred solar masses per year and containing significant dust. These predictions are consistent with the abundance of JWST/HST galaxies selected photometrically in the rest-frame UV \citep{Oesch2018,Bouwens2021,Naidu2022, Kauffmann2022,Donnan2023,Harikane2023,PG2023,sunUltravioletLuminosityFunction2024,sunFaintEndUV2026a,pangGLASSJWSTEarlyRelease2025}, as well with the properties of recently spectroscopically confirmed JWST/HST galaxies \citep{Bouwens2022,Heintz2023,Fujimoto2023,Williams2023,Harikane2023spec,wangStrongHeIi2024,zhouDiscoveryGasenshroudedBroadline2026a} formed during this epoch.

Discrepancies with Labb\'e et al. may arise from overestimation of the stellar masses and the absence of JWST/MIRI data \citep[see][]{Papovich2023}, heavy dust extinction affecting UV luminosities, or misidentification of faint red AGN galaxies at closer redshifts \citep{Labbe2023, Arrabal2023}.

Furthermore, spectroscopic confirmations are now solidifying the distances and properties of many early galaxies. JWST NIRSpec campaigns have secured redshifts for galaxies out to $z\approx 13$–14, removing the ambiguity of photometric estimates. Notably, \citet{carnianiSpectroscopicConfirmationTwo2024} confirmed two luminous galaxies at $z\approx14$ (only $\sim$300 Myr after the Big Bang), demonstrating that such bright systems were indeed in place very early. Their spectra show blue continua with Lyman breaks and no obvious emission lines, indicating stellar populations rather than quasars. The surprising abundance of UV-bright galaxies at $z>10$ is thus real, but these objects do not necessarily overthrow standard models. \citet{xiaoAcceleratedFormationUltramassive2024} conducted a systematic JWST/FRESCO survey of 36 massive galaxies at $5<z<9$ and found no fatal tension with $\Lambda$CDM. In their sample, the stellar masses and number densities of typical massive galaxies align with simulations; only the very most extreme cases (three ultra-massive $M_\ast \gtrsim 10^{11} M_\odot$ systems) would demand unusually high baryon-to-star conversion ($\sim50\%$, still well below 100\%). This greatly mitigates the ``impossible galaxy'' crisis: it appears that standard cosmology can accommodate early massive galaxies with plausible adjustments to galaxy formation physics, rather than requiring new cosmological paradigms.

In this letter, we argue that the current high-redshift JWST/HST results, combined with a realistic galaxy formation model,
provide strong confirmation of the standard $\Lambda$CDM cosmology. The key quantity here is the star formation efficiency (SFE), which is defined as the ratio of stellar mass to the mass of baryons inside halo virial radius: ${\rm SFE} = M_*/f_bM_{\rm halo}$, where $f_b$ is the cosmological fraction of baryons.
As a fundamental prediction our model contradicts the common believe found in the literature that the galaxies in the early universe remains constant \cite[e.g.][]{Tacchella2018,Oesch2018,Harikane2023}. We do find that SFE increases at high redshift and reaches a mean value of 3.5\% at $z=13$. 

In Section~\ref{sec:section2}, we compare our model with the rest-frame UV luminosity function at $z\sim 8-12$ derived from JWST/HST galaxy samples. Individual galaxy properties of spectroscopically-confirmed JWST/HST galaxies are examined in Section~\ref{sec:section3}. We present our $\Lambda$CDM model predictions for the SFE in the early universe in Section~\ref{sec:section4} and conclude in Section~\ref{sec:section5}.

\section{The galaxy UV luminosity function at $\lowercase{z}\sim 8-12$} \label{sec:section2}

The \textsc{Uchuu-UM} galaxy catalogues used in this analysis were generated by applying the \textsc{UniverseMachine} \citep{Behroozi2019} to assign galaxies to the dark matter halos in Shin-Uchuu and Uchuu1000-PL18 $N$-body simulations\footnote{https://www.skiesanduniverses.org/Simulations/Uchuu/} \citep{Ishiyama2021}. 
The \textsc{UniverseMachine} is a self-consistent empirical model that parameterises the galaxy SFR as a function of its host dark matter halo mass, mass accretion rate, and redshift. \textsc{UniverseMachine} was developed to accurately predict realistic global properties for individual high-redshift galaxies as observed by JWST \citep{Behroozi2020}, including UV luminosity, stellar mass, SFR, and dust extinction. Additionally, the model incorporates the evolution in the galaxy-halo relationship. 
Our galaxy catalogues cover redshifts from $z=0$ to 20. We use the same Planck cosmology for both simulations, with particle masses of $8.97 \times 10^5$ $h^{-1}$M$_{\odot}$ for Shin-Uchuu and $3.29 \times 10^8$ $h^{-1}$M$_{\odot}$ for UchuuPL18, and box sizes of 140 $h^{-1}$Mpc and 1000 $h^{-1}$Mpc, respectively.

The original \textsc{UniverseMachine} DR1 catalogues were not trained or tested on any observational data with $z>10$.  As an empirical model, they used a flexible fit to capture the redshift scaling, which resulted in an unphysical shape to the stellar mass -- halo mass relation at $z\ge 10$, with higher-mass halos receiving lower stellar masses than lower-mass halos.  For this paper, we addressed this shortcoming by adding a physical prior that larger-mass halos should always host higher-mass halos (i.e., the stellar mass--halo mass relationship should be monotonically increasing) and refit with the original dataset (i.e., only data with $z\le 10$ without including any JWST data), starting from the previous best-fit model.  We did not re-fit the entire posterior distribution, as this will be performed with JWST data in an upcoming paper (Cooray et al., in prep.).

\begin{figure}[ht!]
\includegraphics[scale=0.31]{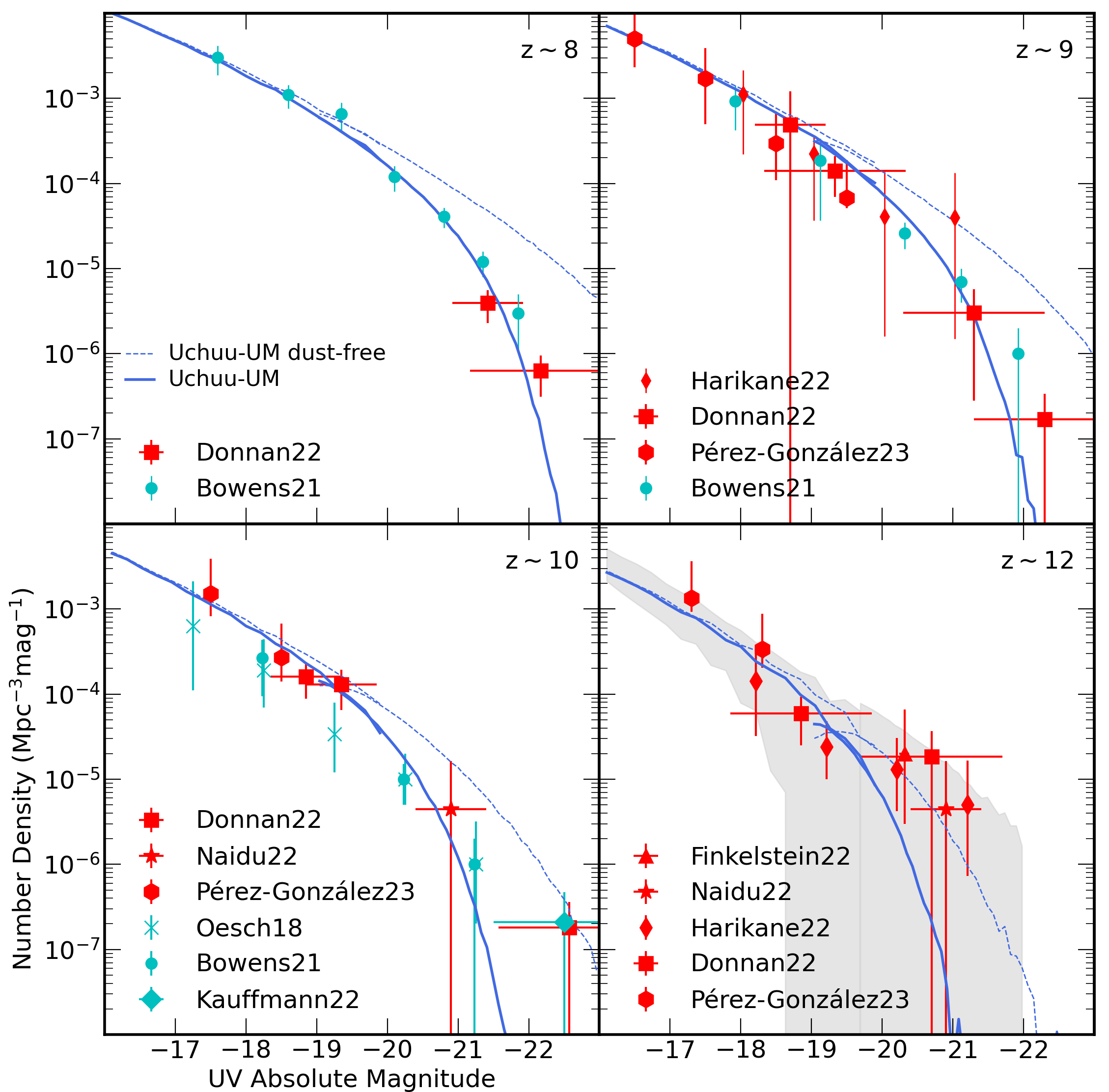}
\caption{Rest-frame UV luminosity functions at $z = 8, 9, 10,$ and $12$. JWST \citep{Naidu2022, Donnan2023, Harikane2023, PG2023, Finkelstein2022} (red) and HST \citep{Oesch2018, Bouwens2021, Kauffmann2022} (cyan) measurements compared to \textsc{Uchuu-UM} predictions with (solid) and without (dotted) dust. The shaded region at $z=12$ indicates cosmic variance.
\label{fig:Figure_1}}
\end{figure}

Figure~\ref{fig:Figure_1} shows the rest-frame UV luminosity function at $z=8$, 9, 10, and 12 derived from galaxy samples selected photometrically in the rest-frame UV. Red symbols represent  measurements from the JWST \citep{Naidu2022, Donnan2023, Harikane2023, PG2023, Finkelstein2022}, while cyan symbols show those from the Hubble Space Telescope (HST)\citep{Oesch2018, Bouwens2021, Kauffmann2022}. We also show theoretical results based on well-tested and widely-used combination of cosmological $N-$body \textsc{Uchuu} simulations \citep{Ishiyama2021} with a galaxy formation algorithm, \textsc{UniverseMachine}, that produces simulated galaxies \citep {Behroozi2019}. Mock galaxy catalogues generated by this approach are called \textsc{Uchuu-UM}. Solid lines indicate the luminosity functions obtained from our \textsc{Uchuu-UM} galaxy catalogues, while dotted lines are for the same galaxies with no internal dust extinction. The shaded region, only shown at $z=12$ as a reference, represents the uncertainty due to cosmic variance considering the observed JWST volume at these redshifts. The \textsc{Uchuu-UM} model accurately reproduces the observed JWST and HST UV luminosity functions over a range of five UV absolute magnitudes.

\begin{figure}[ht!]
\includegraphics[scale=0.47]{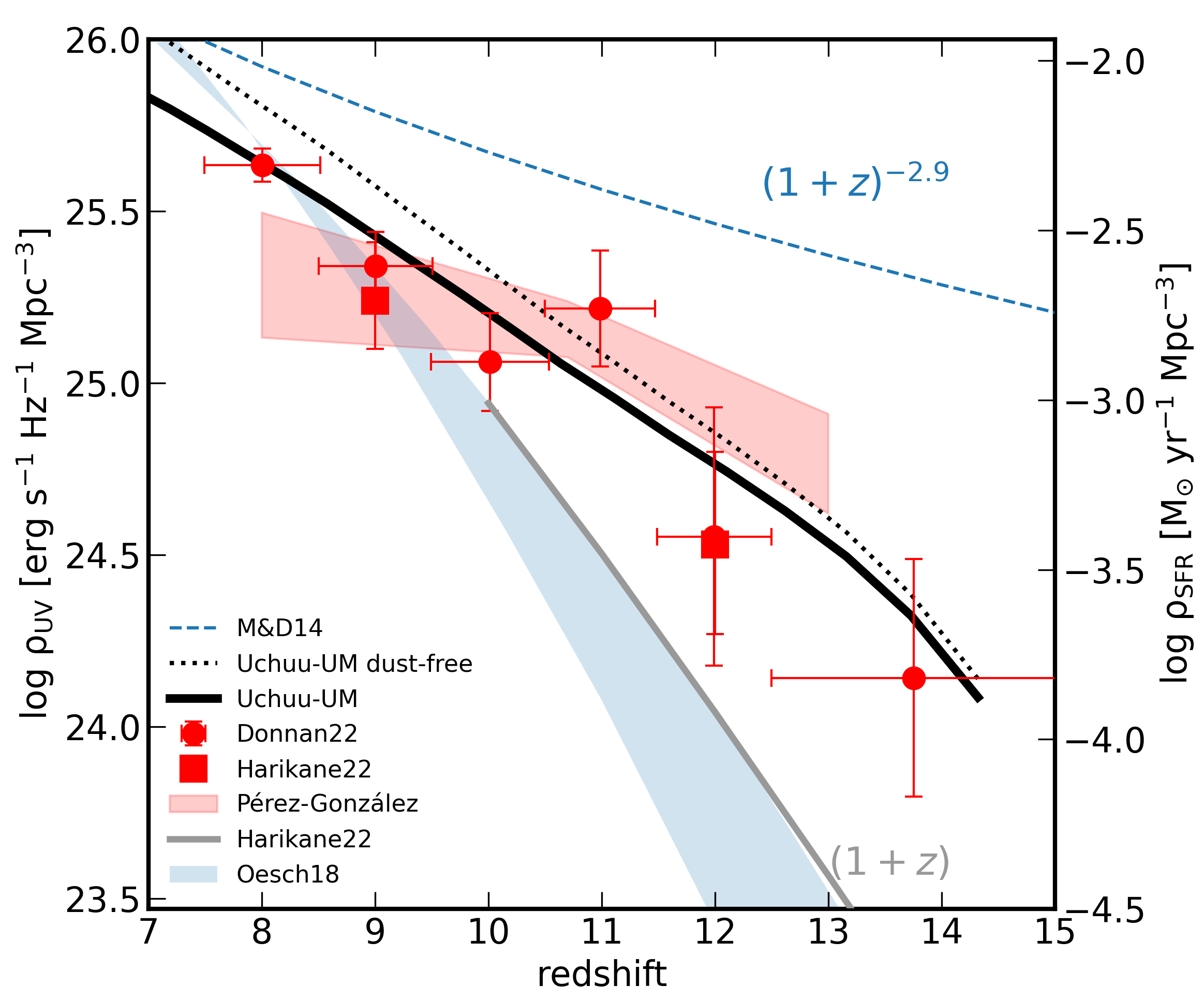}
\caption{Evolution of UV luminosity and SFR densities for galaxies with $M_{\mathrm{UV}} < -17$. JWST data \citep{Donnan2023, Harikane2023, PG2023} (red) compared to \textsc{Uchuu-UM} predictions with (solid) and without (dotted) dust, reproducing the observed evolution over $7 \lesssim z \lesssim 14$.
\\
\\
\label{fig:Figure_2}}
\end{figure}

Figure~\ref{fig:Figure_2} shows the evolution of the cosmic UV luminosity and SFR density in galaxies integrated down to $M_{\rm UV}=-17$. Recent measurements with JWST are denoted by red symbols \citep{Donnan2023, Harikane2023,PG2023}. The solid black line shows the results from  \textsc{Uchuu-UM} catalogues; the dotted line is the same without internal dust extinction. The shaded cyan region depicts the halo evolution model from Oesch et al. \citep{Oesch2018}, which predicts a rapid decline in density at $z > 8$. The extrapolated best-fit by Madau \& Dickinson \citep{MD2014} and the constant star-formation efficiency model from \cite{Harikane2023} are also shown as a reference. Our \textsc{Uchuu-UM} model is in good agreement with the evolution of UV/SFR density determined from JWST over the redshift range $z\sim7$$-$$14$.

\section{Global properties of JWST and \textsc{Uchuu-UM} galaxies} \label{sec:section3}

\begin{figure*}[ht!]
\centering
\includegraphics[scale=0.455]{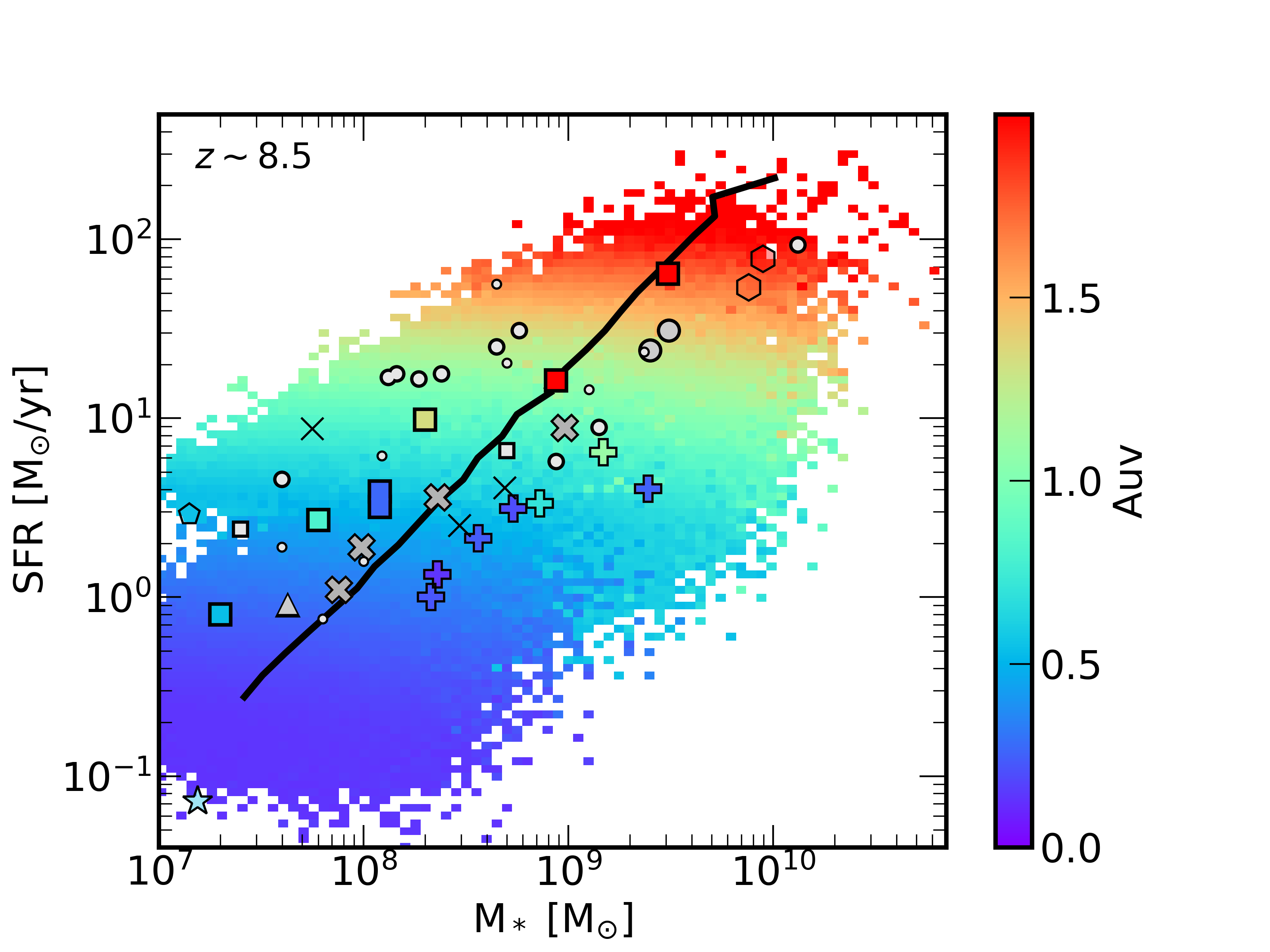}
\includegraphics[scale=0.455]{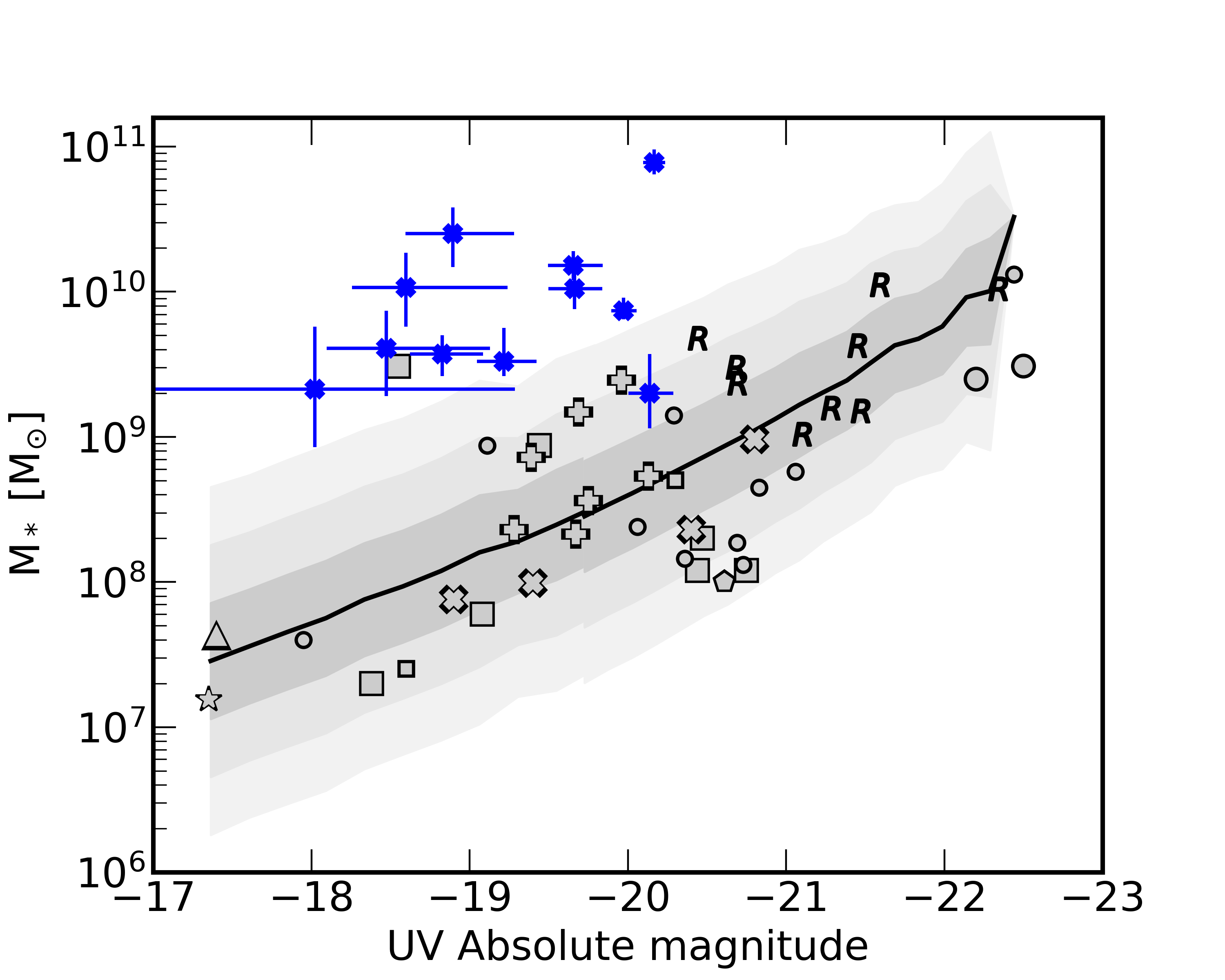}
\caption{\textsc{Uchuu-UM} predictions for spectroscopically confirmed JWST/HST galaxies  at $z\sim8.5$. Symbols are: pentagon for \cite{Heintz2023}, large square for \cite{Fujimoto2023}, large circle for \cite{Bouwens2022}, triangle for \cite{Williams2023}, large plus-cross for \cite{Morishita2023},
hexagon for \cite{Akins2023}, five point star for \cite{Roberts-Borsani2023},
large grey x-cross for \cite{Harikane2023},
small circle for \cite{Nakajima2023},
small square for \cite{Scholtz2025},
smallest circle for \cite{Sarkar2025},
simple x-cross for \cite{Sarkar2025}. \textit{Left:} SFR versus stellar mass, color-coded by $A_{\rm UV}$. \textit{Right:} Stellar mass versus UV luminosity. Observations agree with the model relations, while red candidates from \citet{Labbe2023} (blue crosses) lie significantly above them, with $\sim 50\times$ higher stellar masses at fixed UV luminosity. This discrepancy could be due to overestimation of the stellar masses by a range of effects (systematic uncertainties, lack of JWST/MIRI data \citep{Papovich2023}, heavy dust extinction affecting the UV luminosities, or mistaken red AGN galaxies at closer redshifts \citep{Labbe2023}).
Symbols shape are as in the left figure, plus the points marked 'R' from \cite{Rojas-Ruiz2025}.
\\
\\
\label{fig:Figure_3}}
\end{figure*}

\begin{figure}[ht!]
\centering
\includegraphics[scale=0.42]{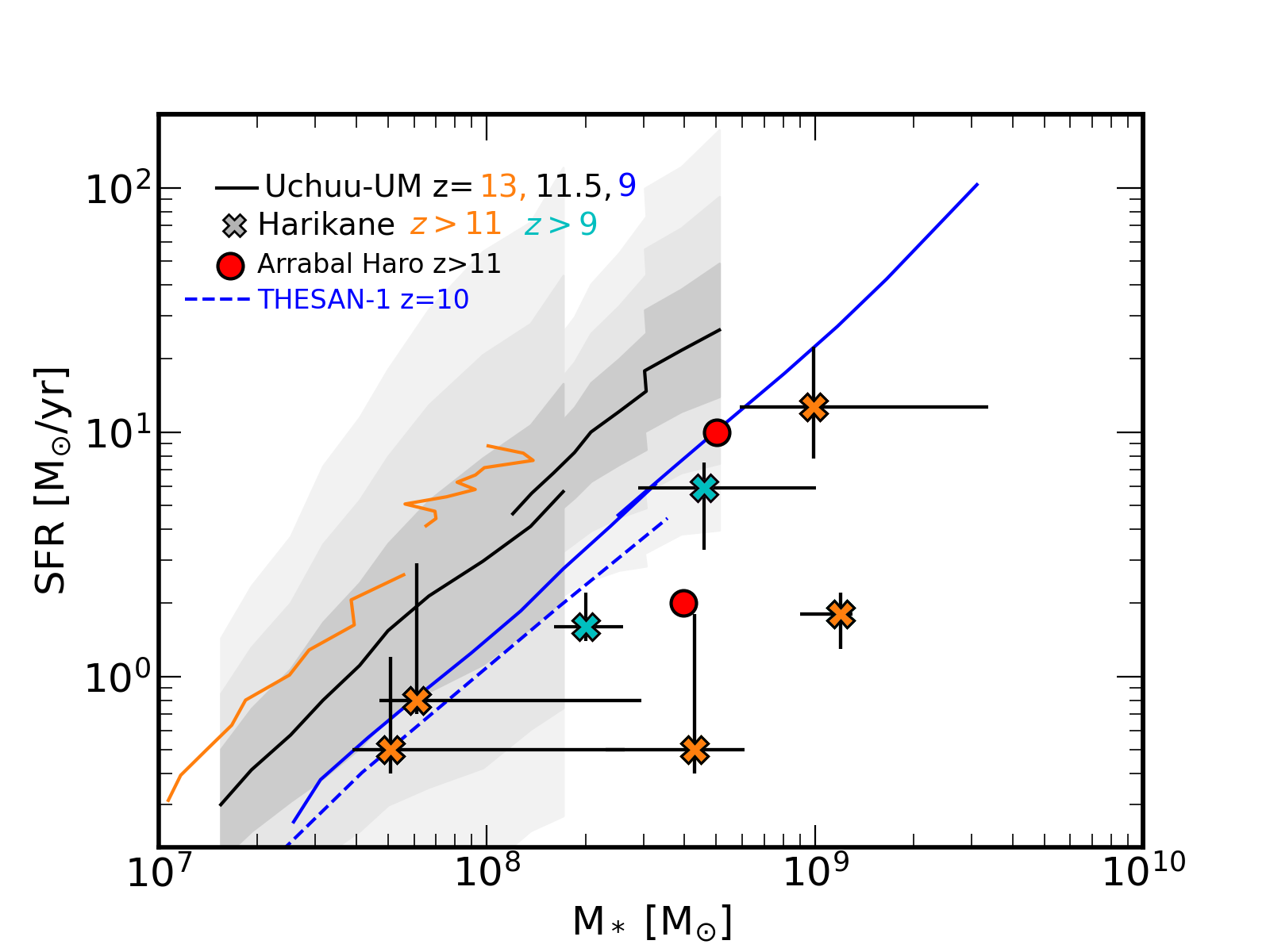}
\includegraphics[scale=0.45]{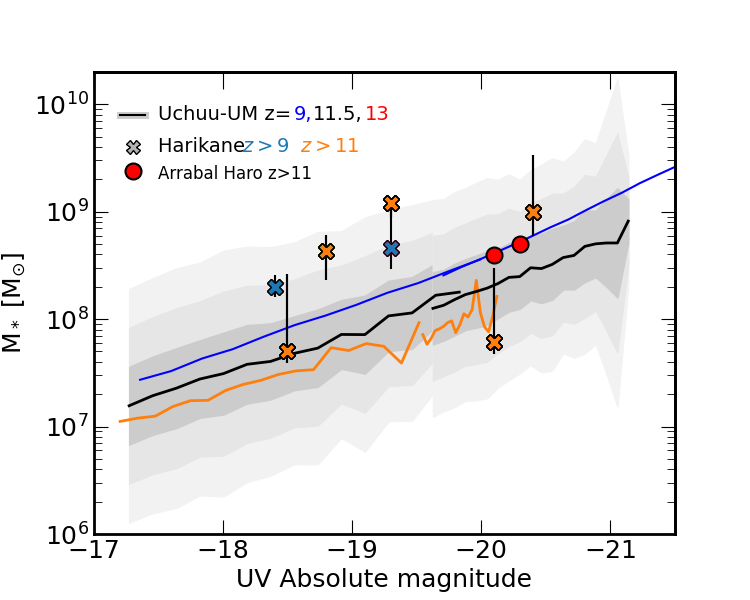}
\caption{Properties of seven spectroscopically confirmed JWST galaxies at $9 \le z \le 13.2$ from \citet{Harikane2023spec,Arrabal2023}. \textit{Top:} SFR versus stellar mass. \textit{Bottom:} Stellar mass versus UV luminosity. Solid lines show the mean relations from \textsc{Uchuu-UM} at the corresponding redshifts. The THESAN-1 hydrodynamic simulation at $z=10$ \citep{Garaldi2023} is in good agreement with \textsc{Uchuu-UM}. At $z>9$, observed galaxies tend to have higher stellar masses at fixed SFR or UV luminosity than predicted, although uncertainties remain large and the sample is limited.
\\
}
\label{fig:Figure_4}
\end{figure}

We validate \textsc{Uchuu-UM} predictions by comparing them with galaxies observed with JWST \citep{Heintz2023,Fujimoto2023,Williams2023} and HST \citep{Bouwens2022}. These galaxies have measured spectra and span the redshift range of $7.7 < z < 9.5$. The properties of these galaxies used in Figures~\ref{fig:Figure_3} and~\ref{fig:Figure_4} are listed in Table 1. Figure~\ref{fig:Figure_3} (left panel) shows the SFR - stellar mass relation of \textsc{Uchuu-UM} galaxies brighter than $M_{\rm UV}=-17$ at $z\sim8.5$ which agrees with the available sample of spectroscopically confirmed JWST/HST galaxies (each symbol is colored according to the corresponding A$_{\rm UV}$ obtained from the galaxy spectra). 

\textsc{Uchuu-UM} also provides a good explanation for the relationship between stellar mass and UV luminosity observed for those galaxies as shown in Figure~\ref{fig:Figure_3} (right). However, the red massive galaxies in Labbé et al., denoted by blue cross symbols, exhibit anomalous properties, with an average stellar mass 50 times greater for their corresponding UV luminosities, resulting in a significantly higher mass-to-light ratio at this epoch. These galaxies show a significant deviation from the $3\sigma$ scatter of the \textsc{Uchuu-UM} galaxies, see Figure~\ref{fig:Figure_3} (right panel). 

Labbé et al. considered the possibility of overestimated fiducial masses by factors of $>10-100$ due to faint red AGN \cite[see][for more details]{Arrabal2023}. Recent findings \citep{Papovich2023} indicate that adding JWST-MIRI data reduces the derived stellar mass by 0.4 dex for most high-redshift galaxies. On the other hand, if we accept their stellar mass estimates, these galaxies likely suffer from significant dust extinction, requiring their intrinsic UV luminosities to be 15-40 times larger than observed, in order to reconcile with the stellar mass - UV luminosity relation shown in Figure~\ref{fig:Figure_3} (right panel).

Figure~\ref{fig:Figure_4} shows the SFR–stellar mass (top) and stellar mass–UV luminosity (bottom) relations for spectroscopically confirmed JWST galaxies at $9 \le z \le 13.2$, as taken from \cite{Harikane2023spec,Arrabal2023}. The solid lines represent the mean relations derived from our \textsc{Uchuu-UM} galaxy catalogs at the corresponding redshifts, as indicated in the legends. Notably, the THESAN-1 radiation-hydrodynamic simulation by \cite{Garaldi2023} at $z=10$ aligns well with {Uchuu-UM}. Our analysis reveals an intriguing observation: JWST galaxies at $z>9$ exhibit higher stellar masses than anticipated by our star formation model for a given SFR/UV luminosity. However, it is important to note the large uncertainties in stellar mass measurements, and the limited sample size warrants caution in drawing definitive conclusions.

\section{Star formation efficiency in the early universe} \label{sec:section4}

\begin{figure}[ht!]
\includegraphics[scale=0.47]{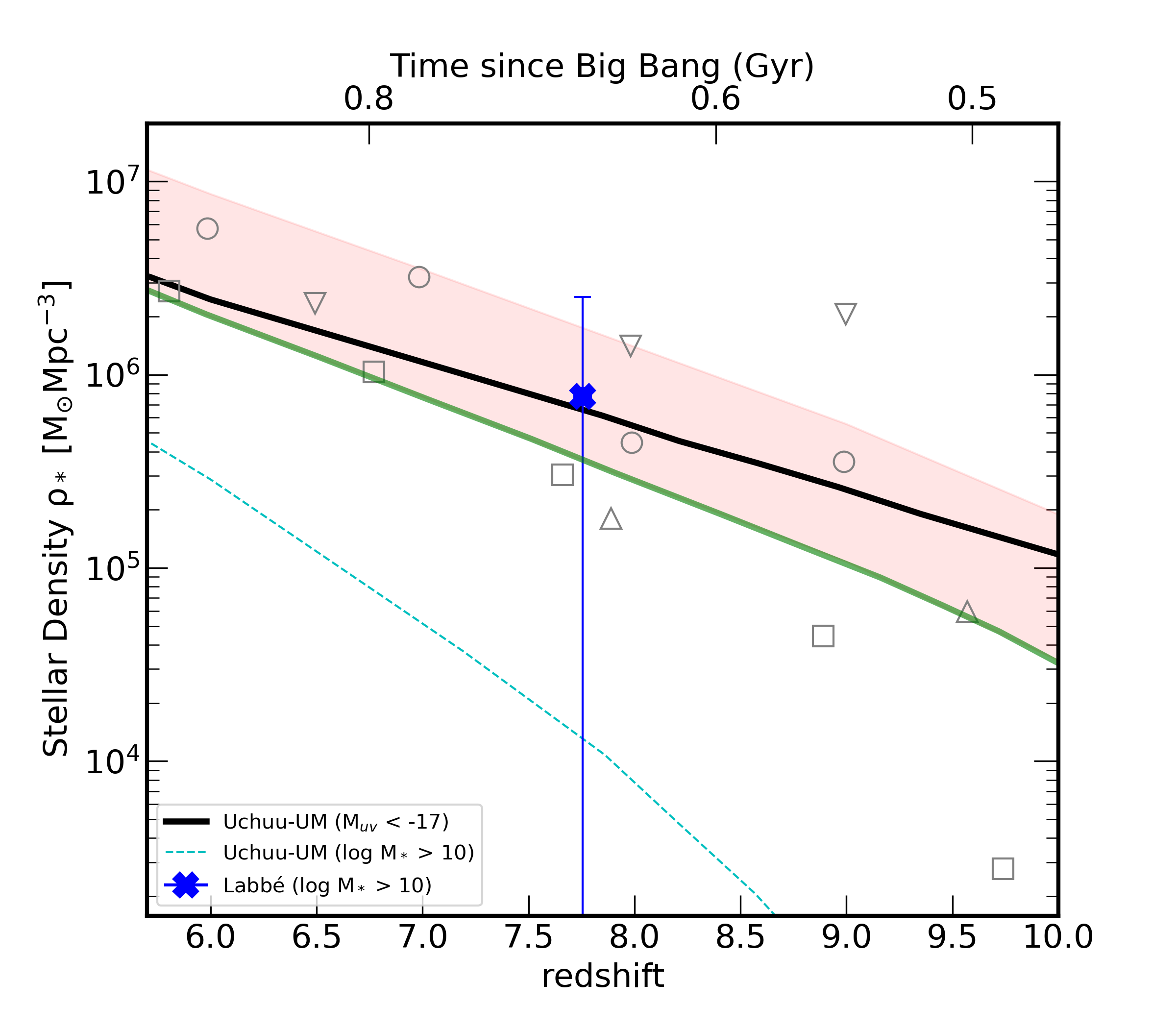}
\caption{Evolution of the cosmic stellar mass density. The \textsc{Uchuu-UM} prediction for galaxies with $M_{\mathrm{UV}} < -17$ (equivalent to a few times $10^7\,M_{\odot}$) is shown by the black curve, in agreement with measurements compiled by \citet{Papovich2023} (gray symbols) and the maximum allowed density (shaded region). The empirical model of \citet{Finkelstein2016}, corrected using JWST/MIRI constraints, is shown in green. For comparison, the estimate of \citet{Labbe2023} for galaxies with $M_* > 10^{10}\,M_{\odot}$ (blue cross) is contrasted with the corresponding \textsc{Uchuu-UM} prediction (cyan dashed line), exceeding it by $\sim 2$ orders of magnitude at $z \sim 8$.
\label{fig:Figure_5}}
\end{figure}

Figure~\ref{fig:Figure_5} shows the evolution of the cosmic stellar mass density, with the black line representing the \textsc{Uchuu-UM} prediction for galaxies brighter than $M_{UV}=-17$ (equivalent to a few times $10^7\,M_{\odot}$) over the redshift range $5.5 < z < 10$. 
This prediction is in agreement with recent measurements compiled by \cite{Papovich2023} (gray open symbols) 
and with the maximally allowable density assuming that galaxies experience a burst at $z=100$ followed by normal star formation (shaded region), as reported in their work. The green line indicates the empirical model of \cite{Finkelstein2016}, corrected by  \cite{Papovich2023} when JWST/MIRI data constrain the stellar masses. Additionally, we compare Labbé et al.'s estimate of the stellar-mass density for galaxies above $10^{10}\,M_{\odot}$ (blue cross, with systematic errors estimated from their Table 2) to the \textsc{Uchuu-UM} prediction for masses above $10^{10}\,M_{\odot}$ (cyan dashed line). Notably, the Labb\'e et al. estimate is two orders of magnitude larger than the expected value at $z\sim8$. 

Labb\'e et al. results, if correct, pose a challenge to the standard cosmology. According to \cite{Boylan-Kolchin2023} they would imply near 100\% efficiency of converting ``normal" gas to stars. Another ``explanation" would be that Labb\'e et al. galaxies are extremely rare events \citep{Lovell2023}. None of those extreme assumptions are required if we assume recent JWST measurements as indicated in Figures~1-4. Indeed, the efficiency of star formation in \textsc{Uchuu-UM} is typical for galaxies such as our Milky Way. There is no need to resort to extreme values statistics either: observed galaxies at high redshifts are typical objects. The strikingly large stellar-mass density observed in galaxies with stellar masses exceeding $10^{10}\,M_{\odot}$ at $z\sim8$, as reported by Labb\'e et al., is a crucial point of conflict with the standard  galaxy formation model.
Our claims presented in this letter add weight to the 
presence of potential systematic errors in Labb\'e et al.

\begin{figure*}[ht!]
\centering
\includegraphics[scale=0.42]{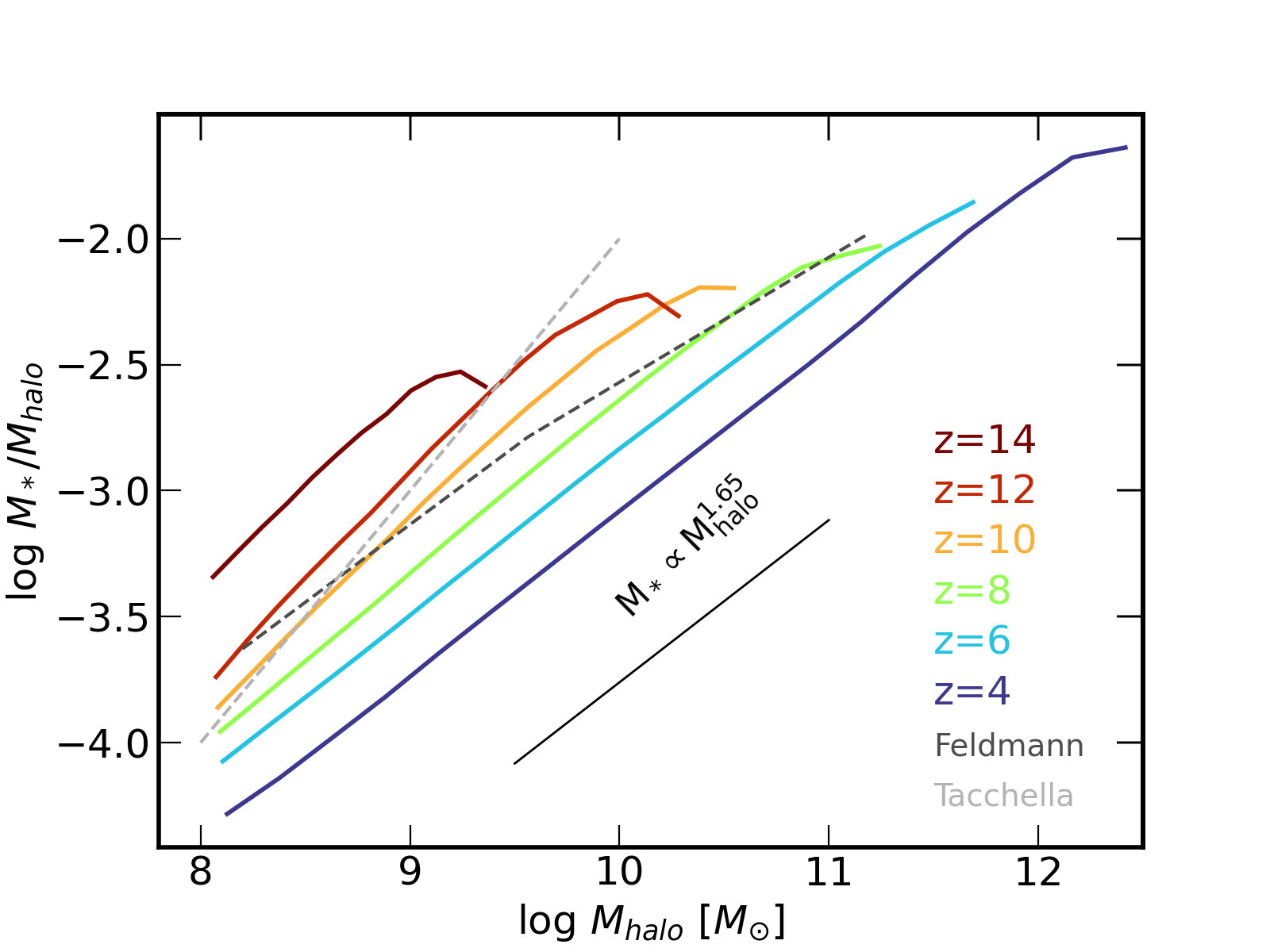}
\includegraphics[scale=0.54]{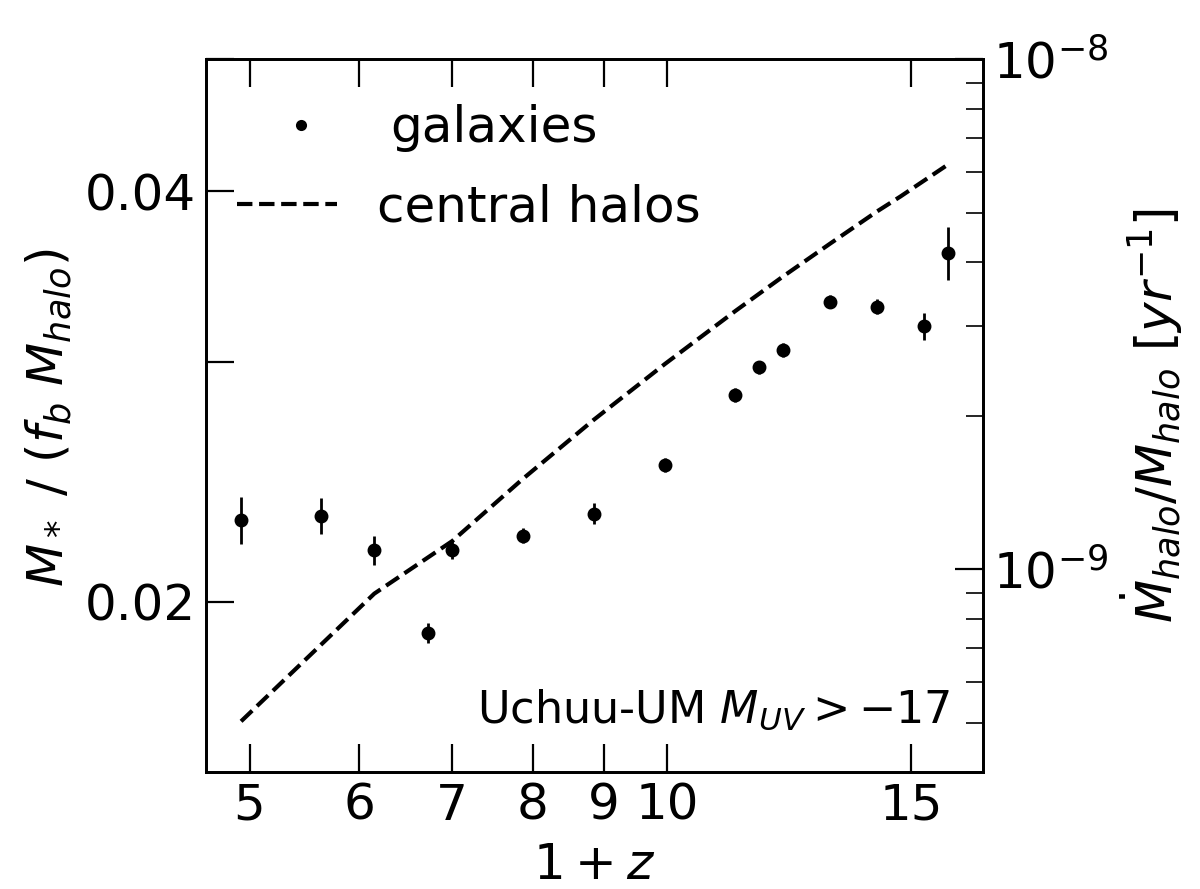}
\caption{Star formation efficiency of \textsc{Uchuu-UM} galaxies. \textit{Left:} Efficiency as a function of halo mass at several redshifts, showing an increase with redshift at fixed halo mass. The dark gray dashed line indicates the FIREbox$^{HR}$ relation \citep{Feldmann2024}, and the light gray dashed line shows $M_* \propto M_{\rm halo}^2$ from \citet{Tacchella2018}; at $z=4$, \textsc{Uchuu-UM} yields $M_* \propto M_{\rm halo}^{1.65}$. \textit{Right:} Redshift evolution of the average star formation efficiency for galaxies with $M_{\mathrm{UV}} < -17$, increasing toward $z>10$, which contradicts the commonly adopted `Constant SFE' model and can explained the observed JWST UV/SFR densities.
\label{fig:Figure_6}}
\end{figure*}

In Figure~\ref{fig:Figure_6}, we present the predicted \textsc{Uchuu-UM} stellar mass-halo mass relation at several selected redshifts greater than four, as indicated by the labels (left panel). The solid curves depict the mean values of the distribution for our simulated \textsc{Uchuu-UM} galaxies. Additionally, we show the results from the MillenniumTNG hydrodynamic simulation \citep{Kannan2023} for comparison (dashed curves), showcasing excellent agreement with \textsc{Uchuu-UM} and demonstrating the reliability of our star formation model in capturing the relevant physics processes involved in galaxy formation within the standard cosmology. 

The star formation efficiency of \textsc{Uchuu-UM} galaxies for the same redshifts, defined as $M_*/(f_b M_{halo})$, as a function of halo mass, are displayed in Figure~\ref{fig:Figure_6} (right panel). The plot clearly shows how the efficiency of star formation increases with redshift. A redshift-independent star formation model (‘Constant SFE’), as that presented in \cite{Tacchella2018} and advocated by \citet{Harikane2024}  to explain the JWST UV/SFR density evolution at $z>8$, significantly under predicts stellar masses. Furthermore, the slope of stellar mass-halo mass relation is much shallower than the $M_* \propto M_{halo}^2$ reported by \cite{Tacchella2018}. We observe a linear relation between the stellar mass and halo mass at $z>5$ (cf. Figure~\ref{fig:Figure_6} left panel) resulting from the galaxy star formation rate in our model being a function of the rate of change of the halo's maximum circular velocity, $V_{max}$ \citep{Behroozi2019,Behroozi2020}.

\section{Discussion and Conclusions} \label{sec:section5}

We show that the abundance and properties of luminous galaxies observed by JWST at $z > 8$ are consistent with expectations from the standard $\Lambda$CDM framework within the assumptions of our \textsc{Uchuu-UniverseMachine} model, assuming that the stellar mass--halo mass relation is monotonically increasing at fixed redshift. The model reproduces the rest-frame UV luminosity functions, the stellar mass--SFR and stellar mass--UV luminosity relations of spectroscopically confirmed galaxies, and the evolution of the cosmic UV and SFR densities over $7 < z < 14$.

A key result is that the star formation efficiency (SFE) increases with redshift at fixed halo mass, reaching $\sim 2$--3\% at $z \sim 10$--12, comparable to the peak efficiency of Milky Way--mass galaxies today. This behavior contrasts with commonly adopted constant-efficiency models \citep[e.g.,][]{Tacchella2018,Harikane2023}, which underpredict both stellar masses and UV luminosities at high redshift. The increase in SFE reflects the high gas accretion rates and compact dynamical scales of early halos, which drive intense but short-lived starbursts. In our model, the linear scaling $M_\ast \propto M_{\rm halo}^{1.65}$ at $z=4$ naturally arises because the instantaneous star formation rate is tied to the time derivative of the halo's maximum circular velocity, $\dot{V}_{\max}$ \citep{Behroozi2019,Behroozi2020}. This result is subject to uncertainties in dust attenuation, star formation histories, and stellar population modeling, but provides a consistent explanation of current observations.

The apparent excess of massive, red galaxies reported by \citet{Labbe2023}  at $z \sim 8$ can be explained without invoking departures from $\Lambda$CDM. Their inferred stellar masses are plausibly affected by systematic uncertainties, including dust attenuation, AGN contamination, and the absence of JWST/MIRI constraints \citep{Papovich2023}. Accounting for these effects brings their number densities into agreement with model predictions and with other recent hydrodynamical simulations \citep[e.g.,][]{Kannan2023,Garaldi2023}.

In summary, our results indicate that (1) the $\Lambda$CDM framework reproduces the population of luminous galaxies observed by JWST up to $z \sim 14$, (2) the star formation efficiency increases with redshift, reaching $\sim 4\%$ at early times, and (3) no exotic physics or extreme baryon conversion efficiencies are required. Future JWST and ALMA observations that constrain stellar ages, dust content, and gas fractions (see, e.g., Cooray et al., in preparation) will be critical to test these predictions and to refine the physical picture of how the first galaxies assembled in the early universe.

\section{Acknowledgments} \label{sec:sectionA}

We acknowledge discussions with E. Alfaro, C-A. Dong-P\'aez, Y. Dubois, Y. Harikane, and M. Volonteri. We thank I. Labb\'e and Y. Harikane for their guidance on the calculation of UV absolute magnitudes. FP, AK, and EP acknowledge support from the Spanish MICINN funding grant PGC2018-101931-B-I00 and Severo Ochoa grant CEX2021-001131-S funded by MCIN/AEI/10.13039/501100011033.
PB was partially funded by a Packard Fellowship, Grant \#2019-69646. 
TI has been supported by IAAR Research Support Program in Chiba University Japan, 
MEXT/JSPS KAKENHI (Grant Number JP19KK0344 and JP21H01122), 
MEXT as ``Program for Promoting Researches on the Supercomputer Fugaku'' (JPMXP1020200109 and JPMXP1020230406), and JICFuS.
The Shin-Uchuu simulation was carried out on the Aterui II supercomputer at CfCA-NAOJ and Uchuu1000-PL18 was generated at the supercomputer Fugaku at the RIKEN Center for Computational Science (Project ID: hp220173 and hp230204). The construction of merger trees of those simulations were partially carried out on XC40 at the Yukawa Institute Computer Facility in Kyoto University.
We thank IAA-CSIC, CESGA, and RedIRIS in Spain for hosting the Uchuu data releases in the \textsc{Skies \& Universes} site for cosmological simulations. The \textsc{UniverseMachine} and \textsc{Uchuu-UM} data analysis have made use of the $skun6$@IAA-CSIC computer facility managed by IAA-CSIC in Spain (MICINN EU-Feder grant EQC2018-004366-P).
The \textsc{Uchuu-UM} galaxy catalogues used in this work are publicly available at the \textsc{Skies \& Universes} site for cosmological simulations:
\url{https://www.skiesanduniverses.org/Simulations/Uchuu/}

\bibliography{main}{}
\bibliographystyle{aasjournal}


\begin{table}
\caption{Summary of values for Figures 3 and 4.}  
\centering                                      
\begin{tabular}{lcrrrrl}                        
\hline\hline                 
name           & z    & log $M_*$ & SFR & M$_{UV}$ & A$_V$ & reference  \\
\hline                                      
S04590         & 8.4959 & 7.15 & 2.9 & -20.61 & 0.2 & \cite{Heintz2023} \\
RX J2129-z95:G2 & 9.5100 & 7.63 & 1.7 & -17.40 & -- & \cite{Williams2023} \\
CEERS1\_3858   & 8.8070 & 8.08 & 3.2 & -20.44 & 0.1 & \cite{Fujimoto2023} \\
CEERS1\_3908   & 8.0050 & 8.30 & 9.8 & -20.47 & 0.5 & \cite{Fujimoto2023} \\
CEERS1\_3910   & 7.9932 & 8.94 & 16.3 & -19.44 & 1.0 & \cite{Fujimoto2023} \\
CEERS1\_6059   & 8.8760 & 8.08 & 3.9 & -20.75 & 0.1 & \cite{Fujimoto2023} \\
CEERS3\_1748   & 7.7690 & 9.49 & 64.3 & -18.55 & 1.8 & \cite{Fujimoto2023} \\
CEERS6\_7603   & 8.8805 & 7.30 & 0.8 & -18.38 & 0.2 & \cite{Fujimoto2023} \\
CEERS6\_7641   & 8.9980 & 7.78 & 2.7 & -19.08 & 0.3 & \cite{Fujimoto2023} \\
REBELS-18      & 7.6750 & 9.49 & 31.0 & -22.5 & -- & \cite{Bouwens2022} \\
REBELS-36      & 7.6770 & 9.40 & 34.0 & -22.2 & -- & \cite{Bouwens2022} \\
YD4            & 7.8758 & 9.17 & 6.46 & -19.69 & 1.10 & \cite{Morishita2023} \\
YD7            & 7.8772 & 9.39 & 4.04 & -19.96 & 0.25 & \cite{Morishita2023} \\
ZD6            & 7.8843 & 8.86 & 3.34 & -19.39 & 0.71 & \cite{Morishita2023} \\
YD8            & 7.8869 & 8.36 & 1.34 & -19.28 & 0.14 & \cite{Morishita2023} \\
ZD2            & 7.8800 & 8.73 & 3.13 & -20.13 & 0.20 & \cite{Morishita2023} \\
ZD3            & 7.8816 & 8.33 & 1.00 & -19.67 & 0.22 & \cite{Morishita2023} \\
GLASSZ8-2      & 7.8831 & 8.56 & 2.13 & -19.75 & 0.24 & \cite{Morishita2023} \\
JD1            & 9.7930 & 7.19 & 0.07 & -17.35 & 0.55 & \cite{Roberts-Borsani2023} \\
CEERS-24       & 8.998 & 9.8e7 & 1.9 & -19.4 & -- & \cite{Harikane2023spec} \\
CEERS-23       & 8.881 & 7.6e7 & 1.1 & -18.9 & -- & \cite{Harikane2023spec} \\ 
CEERS1\_6059    & 8.876 & 9.6e8 & 8.8 & -20.8 & -- & \cite{Harikane2023spec} \\ 
CEERS1\_3858    & 8.807 & 2.3e8 & 3.6 & -20.4 & -- & \cite{Harikane2023spec} \\ 
ERO\_04590 & 8.496 & 7.6 & 4.6 & -17.95 & -- & \cite{Nakajima2023} \\
GLASS\_10000 & 7.881 & 8.16 & 17.8 & -20.36 & -- & \cite{Nakajima2023} \\
GLASS\_100001 & 7.874 & 9.15 & 8.9 & -20.29 & -- & \cite{Nakajima2023} \\
GLASS\_100003 & 7.877 & 8.27 & 16.6 & -20.69 & -- & \cite{Nakajima2023} \\
GLASS\_100005 & 7.879 & 8.38 & 17.8 & -20.06 & -- & \cite{Nakajima2023} \\
CEERS\_01019 & 8.681 & 10.12 & 93.3 & -22.44 & -- & \cite{Nakajima2023} \\
CEERS\_01023 & 7.779 & 8.76 & 30.9 & -21.06 & -- & \cite{Nakajima2023} \\
CEERS\_01027 & 7.825 & 8.12 & 16.9 & -20.73 & -- & \cite{Nakajima2023} \\
CEERS\_01149 & 8.179 & 8.65 & 25.1 & -20.83 & -- & \cite{Nakajima2023} \\
CEERS\_80083 & 8.637 & 8.94 & 5.8 & -19.11 & -- & \cite{Nakajima2023} \\ 
\hline       
\end{tabular}
\label{tab:sample}                             
\end{table}

\end{document}